\documentclass[twocolumn,preprintn/Users/antiasanchezbotana/Box Sync/draft_scn_mgo/PRL_SUBMISSION/ScN_prb_submission_v7.texumbers,floatfix,showpacs,
prb,superscriptaddress]{revtex4-1}
\usepackage{graphicx}
\usepackage{float}
\usepackage{dcolumn} 
\usepackage{bm}
\usepackage{epsfig}
\usepackage{natbib}
\usepackage{amsmath}
\usepackage{mathtools}
\usepackage{listings}
\usepackage{longtable}
\usepackage{csquotes}
\usepackage{gensymb}
\usepackage{xcolor}
\usepackage{hyperref}
\usepackage{comment}
\hypersetup{
     colorlinks   = true,
     citecolor    = blue
}

\newcommand{\out}[1]{}

\begin{document} 

\title{Charge ordering in Ni$^{1+}$/Ni$^{2+}$ nickelates:  La$_4$Ni$_3$O$_8$ and La$_3$Ni$_2$O$_6$}

\author{A. S. Botana}
\affiliation{Materials Science Division, Argonne National Laboratory, Argonne, Illinois 60439, USA}
\author{V. Pardo}
\affiliation{Departamento de Fisica Aplicada, Universidade de Santiago de Compostela, E-15782 Santiago de Compostela, Spain}
\affiliation{Instituto de Investigacions Tecnoloxicas, Universidade de Santiago de Compostela, E-15782 Santiago de Compostela, Spain}
\author{W. E. Pickett}
\address{Department of Physics, University of California Davis,  Davis, California 95616, USA}
\author{M. R. Norman}
\email{norman@anl.gov}
\affiliation{Materials Science Division, Argonne National Laboratory, Argonne, Illinois 60439, USA}

\date{\today}
\begin{abstract}

\textit{Ab initio} calculations allow us to establish a close connection between the Ruddlesden-Popper layered nickelates and cuprates not only in terms of filling of $d$-levels (close to $d^9$) but also because they show Ni$^{1+}$(S=1/2)/Ni$^{2+}$(S=0) stripe ordering.  The insulating charge ordered ground state  is obtained from a combination of structural distortions and magnetic order. 
The Ni$^{2+}$ ions are in a low-spin configuration (S=0) yielding an antiferromagnetic arrangement of Ni$^{1+}$ S=1/2 ions like the long-sought spin-1/2 antiferromagnetic insulator analog of
the cuprate parent materials. The analogy extends further with the main contribution to the bands near the Fermi energy coming from hybridized Ni-d$_{x^2-y^2}$ and O-$p$ states.

\end{abstract}
\maketitle

\begin{figure}
\includegraphics[width=\columnwidth,draft=false]{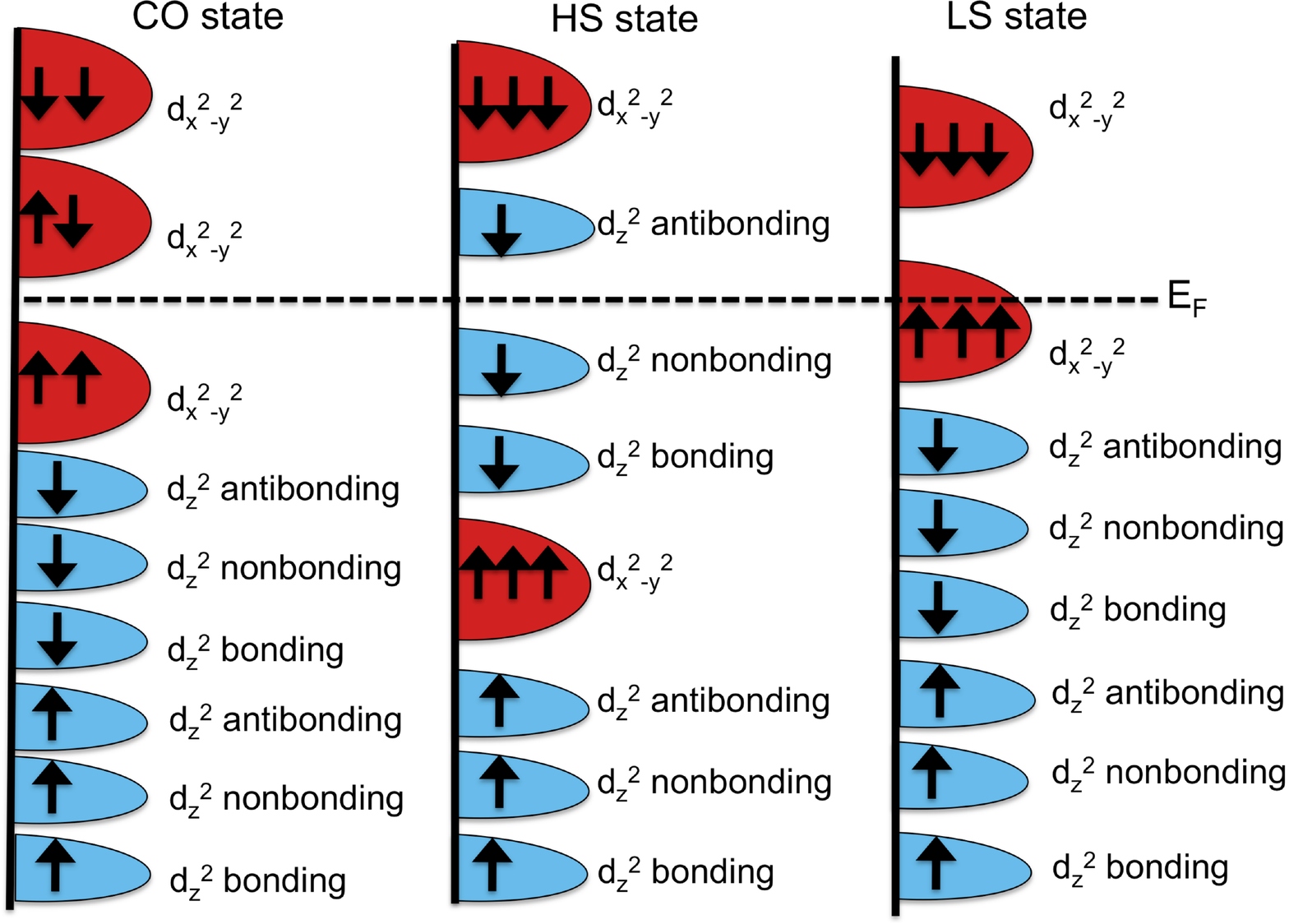}
\caption{(Color online) Level scheme showing the possible spin
states in the trilayer Ni compound La438 (three Ni atoms per f.u.~are represented). On the left side, the purely ionic picture of the low spin charge ordered state - Ni$^{2+}$ (low spin S=0) and 2 Ni$^{1+}$(S=1/2) - gives rise to an insulating state with a gap between d$_{x^2-y^2}$ subbands as in the cuprates. In the center, the high spin state gives rise to an insulating molecular state with a gap between d$_{z^2}$ subbands. On the right, the metallic low spin state is shown. 
}\label{spin_state}
\end{figure}

Layered nickelates have been regarded as the best analog of high temperature superconducting cuprates if the Ni$^{1+}$ state can be stabilized in analogy to Cu$^{2+}$ \cite{CupratelikeFermi}. The discovery of the  Ruddlesden-Popper phases Ln$_{n-1}$(NiO$_2$)$_n$Ln$_2$O$_2$ (Ln=La, Pr, Nd; n=1, 2, 3) with $n$ cuprate-like NiO$_2$ layers reinvigorated the interest in nickelates \cite{Greenblatt1997RP,Zhang_Greenblatt_La3Ni2O7, Greenblatt_SyntheticMetals,Poltavets_La3Ni2O6JACS, Poltavets_Ln4Ni3O8, Poltavets_La3Ni2O6PRL,Poltavets_2010PRL,zhang_arxiv}.  Within this series, the trilayer  La$_4$Ni$_3$O$_8$ (La438) and bilayer La$_3$Ni$_2$O$_6$ (La326) compounds are ionic but highly unconventional insulators \cite{Poltavets_La3Ni2O6PRL,Poltavets_2010PRL}. As the $n$=3 and $n$=2 members of the series, they have a formal Ni valence of +1.33 and +1.5, respectively, which being non-integer should correspond to metallic 
behavior, yet both are insulating. 
The NiO$_2$ slabs are separated by fluorite structure LaO blocking layers that make the inter-trilayer/bilayer coupling very weak. 
The lack of apical oxygen ions reduces the interplane separation substantially and opens a large crystal field splitting for the $e_g$ states
with the d$_{x^2-y^2}$ state lying higher in energy than d$_{z^2}$. The d$_{z^2}$ orbitals hybridize along the $z$ direction giving rise to $n$ molecular subbands. Depending on the relative magnitude of the crystal field splitting and Hund's rule coupling, the ground state can be either high spin (HS) and insulating or low spin (LS) and metallic as depicted in Fig.~\ref{spin_state}.

The $n$=3 La438 compound undergoes a phase transition to an insulating state at 105 K, accompanied by a dramatic increase in the resistivity and a discontinuity in the magnetization \cite{Poltavets_2010PRL, zhang_arxiv}.  NMR experiments reveal the presence of spin fluctuations below 160 K \cite{AprobertsWarren_NMRLa4Ni3O8}. From a theoretical point of view, the insulating character of La438 was accounted for in terms of a molecular high spin (HS) state with the insulator-to-metal transition being spin driven \cite{Pardo_Quantum,Pardo_La3Ni2O7,Pardo_PressureMIT}. With the gap being formed between d$_{z^2}$ bands, the electronic structure of the high spin state differs from the one in cuprates.

Although the trilayer nickelate exhibits a transition likely accompanied by antiferromagnetic (AFM) order, the insulating bilayer material shows no transition down to 4 K 
\cite{Poltavets_La3Ni2O6PRL}. Transport and magnetic measurements have shown that La326
is a paramagnetic insulator with spin fluctuations similar to those seen in La438 \cite{NMR_La3Ni2O6}. 
The $e_g$ crystal field splitting, Hund's rule coupling, and ostensibly the AFM exchange interactions should be comparable to those in the trilayer material. The different electron count (and Ni average valence) leads to the presence of two versus three d$_{z^2}$ orbitals forming the molecular basis. 
Generally viewed, the physics of the spin states and possibilities for an insulating molecular state are quite similar in La326 and La438, though the different energy scales are such that the former has not shown a transition yet in the temperature range studied 
\cite{Poltavets_La3Ni2O6PRL,NMR_La3Ni2O6}.

Recently, Zhang \textit{et al.}~\cite{zhang_arxiv} showed using x-ray diffraction on single crystals of La438 that the transition is associated with real space ordering of charge within each plane forming a striped ground state.
The superlattice propagation vector is oriented
at 45$^{\circ}$ to the Ni-O bonds with the stripes being weakly correlated along $c$ to form a
staggered AB stacking of the trilayers. Within each trilayer, the stripes are stacked in phase from
one layer to the next. This planar charge modulation provides an alternate route to the insulating state as compared to the previous picture based on molecular orbitals formed by hybridization along $c$.

Here, using density functional theory (DFT)-based calculations, we show  that a charge ordered phase of Ni$^{1+}$ (S=1/2)/Ni$^{2+} $(S=0) stripes has a lower free energy than the previously suggested molecular insulating state in La438, and this accounts for both the insulating
nature and the superlattic peaks seen experimentally. The gap opens up from a combination of charge-ordered related structural distortions and exchange splitting and is formed solely within the d$_{x^2-y^2}$ manifold of states. When doped, electrons or holes would go into these states, in a similar fashion to what occurs in the cuprates. Analogous calculations suggest that checkerboard charge order should appear in the bilayer nickelate La$_3$Ni$_2$O$_6$. Confirmation would ideally require single crystals of La326, which are challenging to synthesize, as well as studies at higher temperatures to access the transition itself.

\begin{figure}
\includegraphics[width=\columnwidth,draft=false]{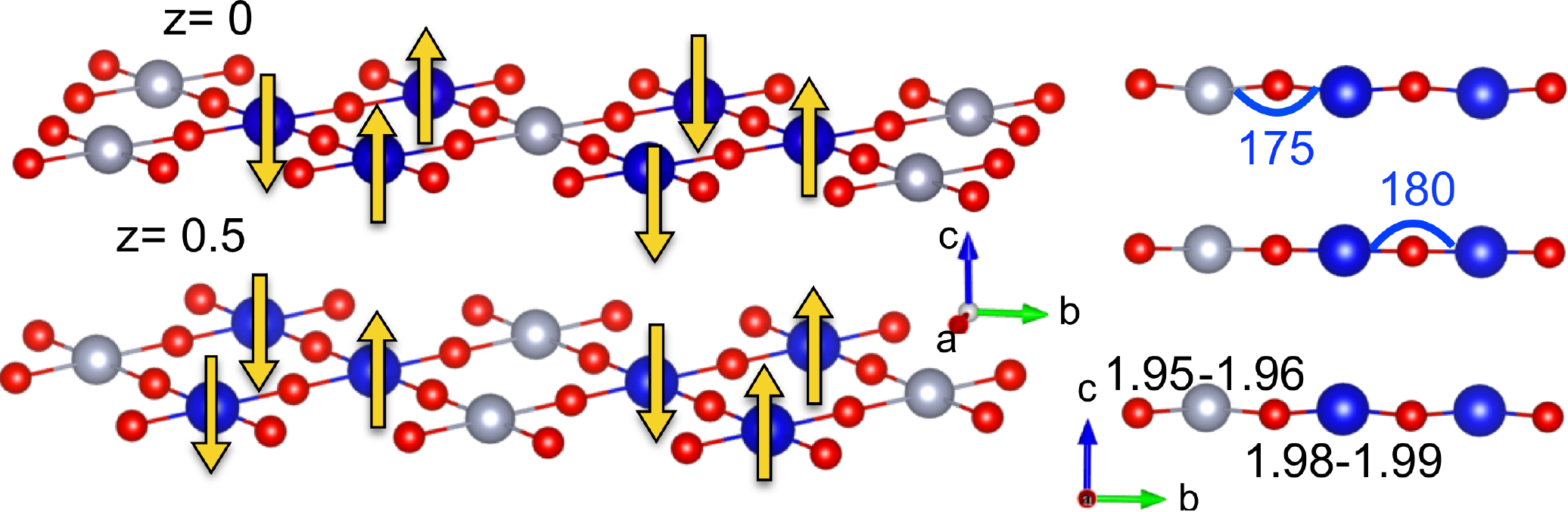}
\caption{(Color online) Left panel: proposed charge/spin ordering pattern inside the NiO$_2$ planes in La$_4$Ni$_3$O$_8$, with Ni$^{1+}$ in blue and Ni$^{2+}$ in gray. Arrows correspond to up/down spins for the Ni$^{1+}$ ions. Right panel: structure after relaxation showing the buckling in the outer layers and the Ni-O bond lengths.}\label{fig1}
\end{figure}

DFT calculations were performed using the all-electron, full potential code WIEN2K \cite{wien2k} based on an augmented plane wave plus local orbital (APW+lo) basis set \cite{sjo}, with atomic positions taken from a recent crystal structure refinement \cite{zhang_arxiv}. For the structural
relaxations, we have used the Perdew-Burke-Ernzerhof
version of the generalized gradient approximation (GGA) \cite{pbe}. 

Our charge-ordered ground state configuration for La438 is found even in the absence of an on-site Coulomb
repulsion $U$ and Hund's rule coupling strength J$_H$. But, to compute more reliably the total energy difference between the 2D striped phase and the 3D molecular insulating state, the
LDA+$U$ scheme has been applied using the so-called fully localized
version for the double-counting correction \cite{sic,fll} that incorporates a $U$ and J$_H$ for the
Ni $3d$ states. Chosen values for $U$ and $J_H$ are 4.75 and 0.68 eV, respectively, as used in earlier work \cite{Pardo_Quantum}. Calculations confirm that a 2D charge ordered state is more stable than the previously proposed HS state by 0.4 eV/Ni within LDA+$U$, so large that the particular choice of $U$ is not critical.

\textit{Charge ordering in La438}. In La438, the average formal Ni valence is +1.33. One possibility would be an in-trimer charge-ordered configuration with the outer Ni atoms being Ni$^{1+}$ and the inner Ni$^{2+}$ but this has been shown to be very unfavorable in energy \cite{wu_co}. If all the Ni ions have the same valence, the e$_g$ states, with 2.67 electrons per Ni on average 
can occur in two different ways: the LS state and the HS state that give rise to a metallic and an insulating state, respectively (Fig.~\ref{spin_state}) \cite{Pardo_Quantum}. These have been the possibilities explored so far, without accounting for any potential in-plane charge ordering that we explore here.

\begin{figure}
\includegraphics[width=\columnwidth,draft=false]{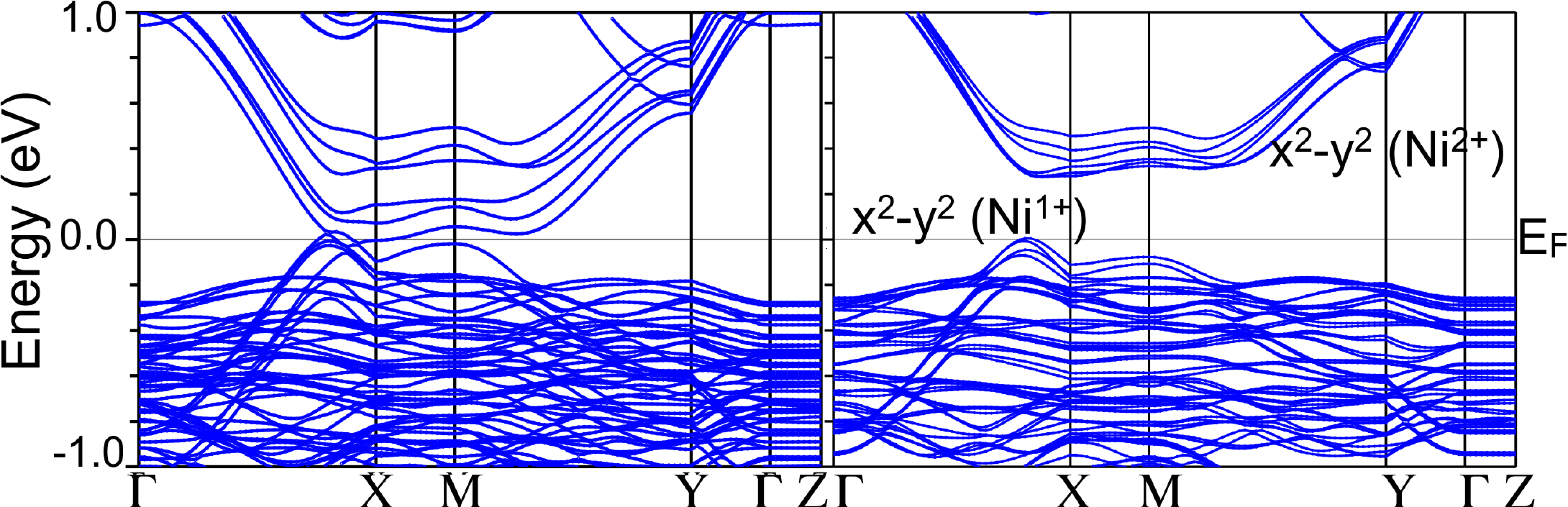}
\caption{(Color online) Band structure of La438 obtained within GGA. Left panel: unrelaxed structure. Right panel: relaxed structure. The gap opening comes about due to structural distortions and is formed solely within the d$_{x^2-y^2}$ manifold of states. Note that both panels are for the charge-ordered supercell.}\label{bs}
\end{figure}

\begin{figure*}
\includegraphics[width=2.1\columnwidth,draft=false]{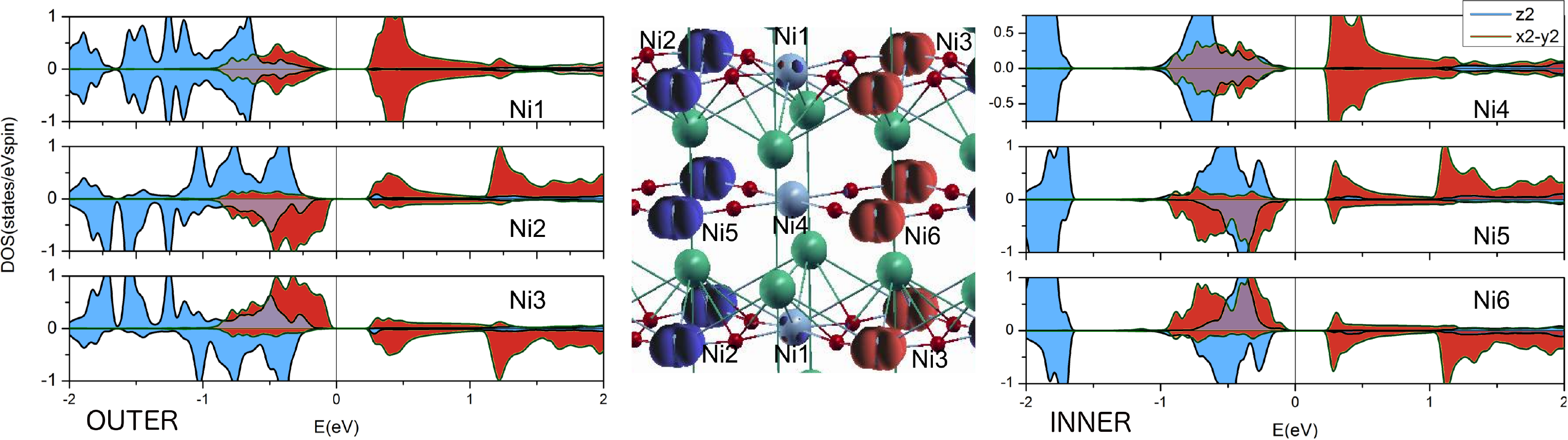}
\caption{(Color online) Calculated orbital resolved $e_g$ density of states for Ni atoms in the trilayer for La438 obtained within GGA. Top curves spin-up, bottom curves spin-down. Left panel: Ni atoms in the outer layers (Ni1, Ni2 and Ni3, as shown in the central panel). Right panel: Ni atoms in the inner layer (Ni4, Ni5 and Ni6, as shown in the central panel). The central panel shows a three-dimensional plot of the spin density in the striped ground state, with an isosurface at 0.1 e/\AA$^3$ obtained using XCrysDen \cite{xcrysden}. Ni1 and Ni4 are the nonmagnetic Ni$^{2+}$ ions. Different colors (shades of gray)
represent the spin-up (spin-down) density.}\label{fig3}
\end{figure*}

To test the possibility of 2D charge ordering, a 3$\sqrt2$$a$ $\times$ $\sqrt2$$a$ $\times$ $c$ supercell was used with the charge/spin pattern shown in Fig.~\ref{fig1}. Formal Ni$^{1+}$:d$^9$ (S=1/2) and Ni$^{2+}$:d$^8$ (S=0) ions in a 2:1 ratio form stripes at 45$^{\circ}$ to the Ni-O bonds with the Ni$^{1+}$ stripes coupled antiferromagnetically. 
Such type of charge and spin ordering yields an AFM arrangement of Ni$^{1+}$ S=$\frac{1}{2}$ ions analog of
the cuprate parent materials. 
Note that the imposed stacking of trilayers is AA and not the experimental AB one since that would require larger supercells. In either case, the coupling between trilayers is weak and within each trilayer the stripes are stacked in phase.

The structure has been relaxed with the lattice constants fixed to the experimental values, so only internal atomic positions were optimized. There is a significant distortion of the Ni-O distances in the NiO$_2$ planes consisting of a modulation of the Ni-O bond length: shorter around the Ni$^{2+}$ ions ($\sim$1.95-1.96 \AA) and longer around the Ni$^{1+}$ ions ($\sim$1.98-1.99 \AA) keeping the average distance very similar to the experimentally reported value. Also, as shown in Fig.~\ref{fig1}, there is significant buckling of the outer NiO$_2$ planes with the inner plane remaining flat. The Ni-Ni distance (both in plane and out of plane) remains unaltered after the relaxation (3.96 \AA~in plane, 3.25 \AA~out of plane) given the fixed lattice constants. 

Charge order-related structural distortions are responsible for the opening of a gap and the corresponding stabilization of the striped phase.  Without distortions, a gap cannot be opened up to the highest $U$ value reached in our calculations of 6 eV. Figure~\ref{bs} shows the band structure for the unrelaxed structure (metallic, on the left) and for the distorted structure after relaxation (insulating, on the right). The insulating character of the derived distorted structure can be observed with a gap of 0.25 eV that opens up near $X$ even without introducing a Coulomb $U$. For a $U$= 4.75 eV the gap increases to only 0.6 eV.

As in cuprates, the gap is of d$_{x^2-y^2}$-only character. From the simple ionic picture, the Ni$^{2+}$ d$^8$ (S=0) cations have two empty d$_{x^2-y^2}$ bands. 
The  Ni$^{1+}$ d$^9$ (S=1/2) ions have one hole in the minority-spin d$_{x^2-y^2}$ band, with the gap being formed between occupied and unoccupied d$_{x^2-y^2}$ states. 
Since the Hund's rule coupling is larger than the bandwidth, the introduction of $U$ is not necessary to open a gap.

To further analyze the electronic structure, Fig.~\ref{fig3} shows the calculated orbital resolved $e_g$ density of states (DOS) of the different Ni atoms. In the striped phase, the Ni$^{2+}$ (d$^8$ LS, S=0) ions have all the $z^2$ bands (majority and minority spin) occupied with the wide d$_{x^2-y^2}$ band for both spin channels remaining unoccupied. For the Ni$^{1+}$ (d$^9$, S=1/2) ions, the d$_{z^2}$ states are also fully occupied and the d$_{x^2-y^2}$ of the minority spin channel is unoccupied.
The DOS clearly shows how the gap is formed between d$_{x^2-y^2}$ bands with predominantly Ni$^{1+}$ character at the top of the valence band and d$_{x^2-y^2}$ bands with predominantly Ni$^{2+}$ character at the bottom of the conduction band. The spin density, pictured in Fig.~\ref{fig3}, also reveals d$_{x^2-y^2}$-only  character. The analogy
with cuprates extends further since there is a high degree of hybridization between Ni-d$_{x^2-y^2}$ and O-$p$ states in the vicinity of the Fermi level. This contrasts with the previously proposed insulating HS state where only d$_{z^2}$ states lie close to the Fermi level and O-$p$ bands are at much lower energies, around 2 eV below the Fermi energy (see Fig.~1 in the Supplementary Material).

Since the discovery of stripe order in high T$_c$ layered cuprates \cite{stripes_cuprates, stripes_cuprates_2}, spin/charge ordering has attracted considerable interest. Our results suggest that the underlying physics of stripe phases in nickelates and cuprates is intimately related in terms of pure electron count and because the stripe ordering of charges and magnetic moments involves bands of d$_{x^2-y^2}$-only character that are highly hybridized with O-$p$ states. Let us recall that similar stripe ordering has been observed and well studied in single layer nickelates, i.e.~La$_{2-x}$Sr$_x$NiO$_4$ (LSNO) \cite{lsno_1,lsno_2,lsno_3}. However, they are further from cuprates in terms of electron count (between d$^{7}$ and  d$^{8}$), spin state, and due to the role that d$_{z^2}$ orbitals play in the vicinity of the Fermi level.

The calculated magnetic moments confirm the formal charge states we have quoted. For the Ni$^{1+}$ ions, the magnetic moments inside the muffin-tin sphere are $\pm$ 0.6-0.7 $\mu_B$. For Ni$^{2+}$, the moment is zero.
To assess the physical Ni charge distributions, the decomposed radial charge densities inside Ni$^{1+}$ and Ni$^{2+}$  spheres were compared directly.
The $3d$ occupations, obtained from the maximum in the radial charge density plots, are
identical for Ni$^{2+}$ and Ni$^{1+}$. The majority and
minority spin valence radial charge densities do differ 
as they must to give the moment, but the
total $3d$ occupation does not differ (see Fig.~2 in the Supplementary Material). This invariance of the actual $d$ electron
occupation (i.e., the charge) in many charge-ordered oxide systems has been discussed in the past \cite{co_1, co_2}. The formal charge of a cation involves the environment of the cation, including the distance to neighboring oxygen ions and the Madelung potentials from the structure (note that the energy difference of the Ni-2$s$ core levels for Ni$^{1+}$ and Ni$^{2+}$ ions is 0.2 eV).   Remarkably, despite the almost equal charge of the two types of Ni atoms, the band structure has a pronounced ionic character reflective of 1+ and 2+ valences.

\begin{figure}
\includegraphics[width=\columnwidth,draft=false]{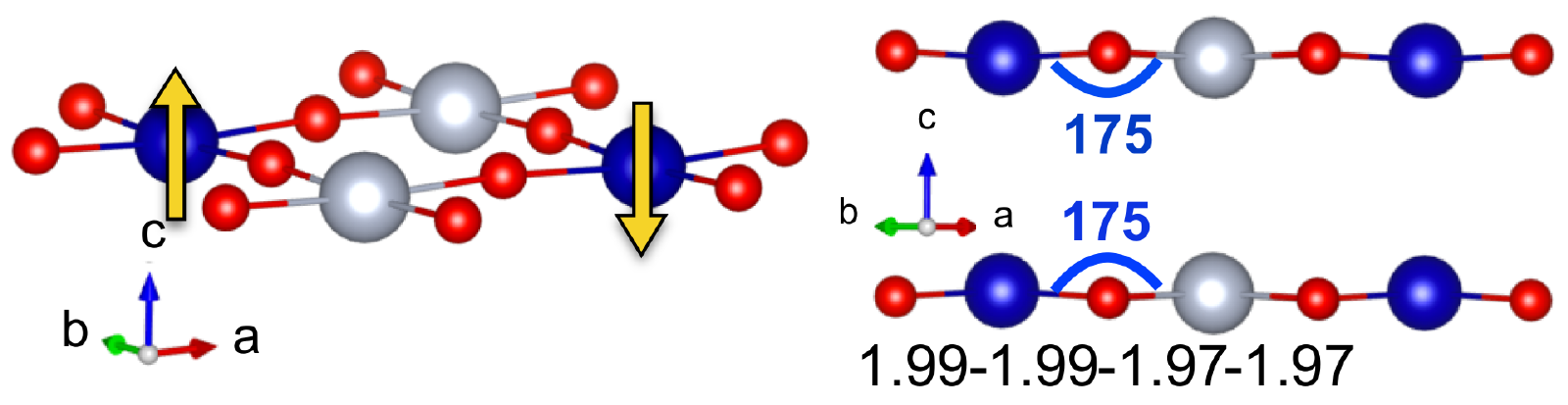}
\caption{(Color online) Left panel: proposed charge and spin ordering pattern in the NiO$_2$ planes for La326, with Ni$^{1+}$ in blue, Ni$^{2+}$ in gray. Arrows correspond to up/down spins for the Ni$^{1+}$ ions. Right panel: Structure after relaxation showing the Ni-O bond lengths in plane  and the buckling of the NiO$_2$ layers.}\label{s326}
\end{figure}

\textit{Charge ordering in La326}. The similarities between La438 and La326 led us to study the possibility of charge ordering in the $n$=2 compound, with an average formal Ni valence +1.5. 
We predict a closely related checkerboard charge ordered insulating phase for La$_3$Ni$_2$O$_6$ 
that provides a 2D AFM spin-half
insulator based on Ni$^{1+}$ (see Fig.~\ref{s326}). The checkerboard phase is more stable than the previously proposed HS state by 0.7 eV/Ni within LDA+$U$ ($U$= 4.75 eV) (again a large energy difference). The structural relaxations within GGA for this magnetic order share the main features with those for La438: a shorter Ni-O bond length for Ni$^{2+}$ atoms ($\sim$1.97 \AA) and a longer one around the Ni$^{1+}$ ones ($\sim$1.99 \AA), as well as significant buckling of the NiO$_2$ plane. 
The magnetic moments obtained are consistent with the Ni$^{2+}$-Ni$^{1+}$ charge ordering picture. For the Ni$^{1+}$ ions, the magnetic moments inside the muffin-tin sphere are $\pm$0.7 $\mu_B$.
For Ni$^{2+}$, the moment is zero.

\begin{figure}
\includegraphics[width=\columnwidth,draft=false]{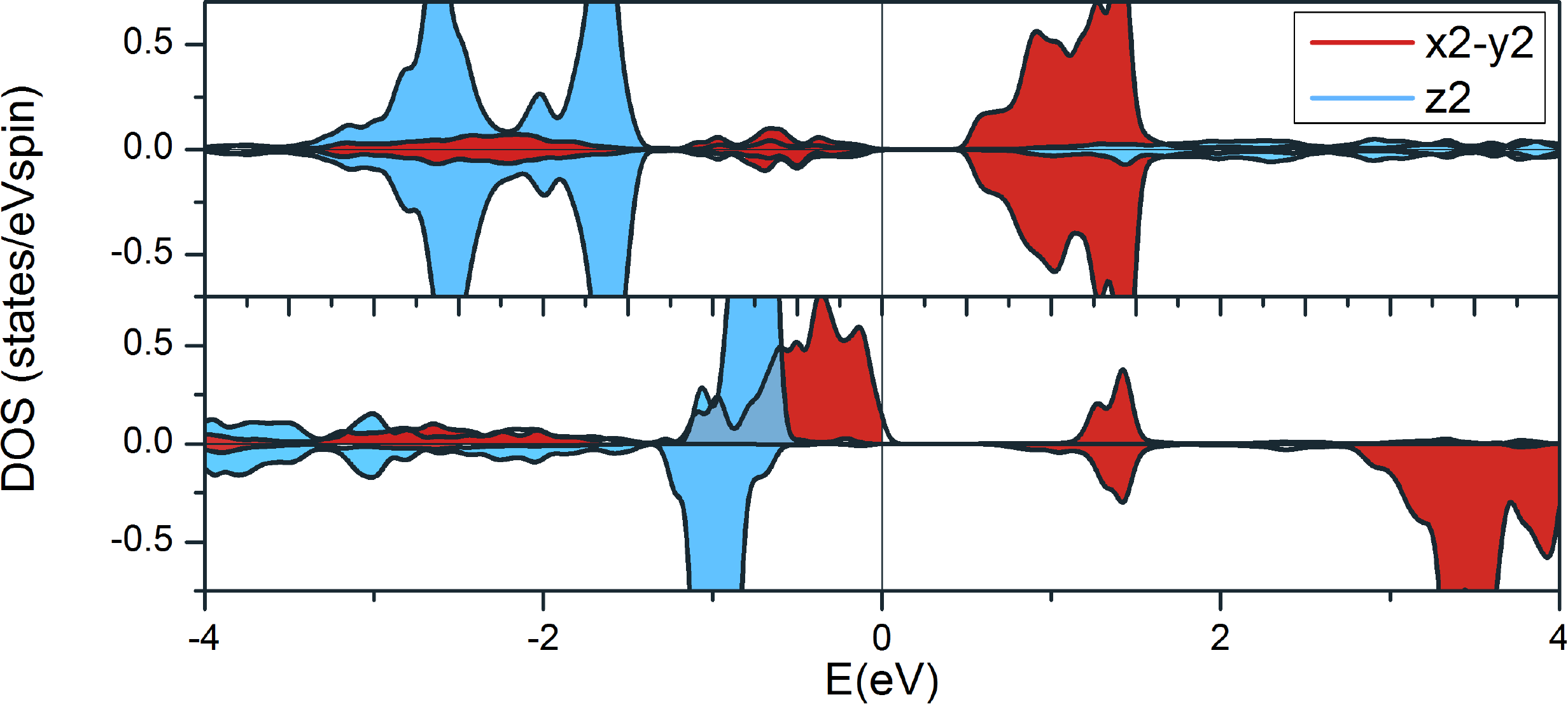}
\caption{(Color online) Calculated orbital resolved $e_g$ density of states for La326 obtained within LDA+$U$ ($U$= 4.75 eV). Top panel: Ni$^{2+}$ atoms. Bottom panel: Ni$^{1+}$ atoms. Top and bottom curves in each panel are majority and minority spin, respectively. Ni$^{2+}$ is non-magnetic. The Ni$^{1+}$ d$_{x^2-y^2}$ states are separated by the Mott-Hubbard gap, $U$.} \label{fig4}
\end{figure}

In the case of La326, the introduction of a $U$ is needed to open a gap in the charge ordered state. Within GGA, the d$_{x^2-y^2}$ bands are wider (W$\sim$2 eV) than for La438. With the gap being formed only between d$_{x^2-y^2}$ states in the charge ordered state, if the Hund's rule coupling is smaller than the bandwidth, a $U$ is needed to obtain an insulating solution. The corresponding orbital resolved density of states for Ni$^{1+}$ and Ni$^{2+}$ in La326  is shown in Fig.~\ref{fig4}. A decoupling of the states is clear: the $e_g$ bands are either Ni$^{1+}$ or Ni$^{2+}$ with negligible mixing and a gap of 0.55 eV. Both Ni sites have all $z^2$ states occupied  with the relatively broad Ni$^{2+}$ d$_{x^2-y^2}$ states unoccupied. The Ni$^{1+}$ ions have d$_{x^2-y^2}$ orbitals split into upper and lower Hubbard bands by the Hubbard $U$, providing a moment near that expected for S=$\frac{1}{2}$. As in La438, the gap is between occupied Ni$^{1+}$-d$_{x^2-y^2}$ and unoccupied Ni$^{2+}$-d$_{x^2-y^2}$ states.

To summarize, \textit{ab initio} calculations give rise to a 2D AFM spin-half
insulating ground state based on Ni$^{1+}$ (pseudo Cu$^{2+}$) stripes for Ruddlesden-Popper layered nickelates. The gap opens from a combination of charge ordered related structural distortions and magnetic order for both La$_4$Ni$_3$O$_8$  and  La$_3$Ni$_2$O$_6$. Our results show a similar electronic structure of these layered nickelates to cuprates not only by pure electron count (close to a d$^9$ configuration), but also because the bands involved around the Fermi level are of of $x^2-y^2$ character only.  These results bring renewed justification that layered nickelates of this type are the cuprate analog systems that are promising for studying the interplay between structure, magnetism, and possibly superconductivity.   

We thank John Mitchell, Junjie Zhang, and Daniel Khomskii for stimulating discussions. Work at Argonne was supported by the Materials Sciences and Engineering
Division, Basic Energy Sciences, Office of Science, US DOE. V.P. thanks MINECO for project MAT2013-44673-R, the
Xunta de Galicia through project EM2013/037, and the Spanish
Government through the Ramon
y Cajal Program (RYC-2011-09024). W.E.P. was supported by Department of Energy Grant No.~DE-FG02-04ER46111.

\end{document}